# The structural and electronic properties of Stone-Wales defective zigzag/armchair antimonene nanotubes: First-principles calculations

Bin Huang[1], Changpeng Chen[1,2,*], Jiaxin Wu[1], Hao Wen[1], Hongzhen Shi[1]
1.School of Science, Wuhan University of Technology, Wuhan 430070, P. R. China
2.Research Center of Materials Genome Engineering, Wuhan University of Technology, Wuhan 430070, P. R. China

**Abstract**
    Geometric optimization and electronic properties of Stone-Wales defective antimonene nanotubes are calculated by the method of first -principle calculations based on density functional theory. Various nanotubes are investigated according to the possible orientations of zigzag/armchair nanostructures when Stone-Wales defects are formed. The band structures, partial density of states and atomic orbitals are calculated to reveal the mechanism of influence of Stone-Wales defects on antimonene nanotubes. When the structure of antimonene changes from monolayer to tube, the indirect gap semiconductor antimonene transforms to a direct gap one. Moreover, the character of direct band gap for the antimonene nanotube is preserved with the Stone-Wales defect forming, while the energy of conduction band bottoms change duo to the intervene of the defect energy level in the band gaps. These results may provide valuable references to the development and design of novel nanotubes based on antimonene nanotubes.
**Keywords**: Stone-Wales defect; antimonene; nanotube; electronic properties; First-principles calculations.

## 1. Introduction

Stimulated by the amazing properties of graphene, scientists have shown great interest in other two dimensional (2D) monolayer materials. Silicene, hexagonal boron nitride, molybdenum disulfide, phosphorene and germanene et al have received increasing attention owing to their unique properties and promising applications[1-19]. Recently, a new ultrathin 2D semiconductor materials in group-V, namely, antimonene, has been grown by van der Waals epitaxy[20].

Defects such as vacancies, adsorption and topological defects, existing inevitably in the fabrication and processing of the two dimensional monolayer materials, may affect dramatically the physical and chemical properties of nanomaterials[1-11]. Stone-Wales (SW) defect being a typical defect which is comprised of two pairs of five-membered and seven-membered rings formed by rotating one bond of the traditional six-membered ring by 90°[21] , have been extensively investigated in graphene and carbon nanotubes [22-27]. While to the best of our knowledge, there are still no investigations on the structural and electronic properties of antimonene nanotube with SW defects.

In this paper, the geometrical structures as well as electronic property of both pristine and SW defective antimonene nanotube are investigated using the density functional theory calculations. Four possible structural defects are considered:

SW1-ZSbNT; SW2-ZSbNT; SW1-ASbNT; SW2-ASbNT. The formation energy of a SW defect are calculated and compared with that of the SW defect in silicene, indicating excellent stability. It is found that the characters of direct band gap for the antimonene nanotubes are preserved with the Stone-Wales defect forming, while the energy of conduction band bottoms change duo to the intervene of the defect energy level in the band gaps.

## 2. Computational methods

The models for the zigzag/armchair antimonene nanotubes(SbNTs) are established(shown in Fig.1 (a)(b)), under the reference of V. Nagarajan's research[28]. There are placed in the center of a cubic supercell. The diameters of zigzag antimonene nanotubes(ZSbNTs) and armchair antimonene nanotubes(ASbNTs) are 17.742Å and 15.579Å, respectively. In order to eliminate the interaction between the adjacent images, the supercell lengths in the plane perpendicular to the direction in which the nanotubes extend are set as 19.031Å and 14.420Å. The side length of the supercell is consistent with the length of the nanotube. The electronic properties and structural optimization of antimonene nanotubes are carried out by using the density functional theory method(DFT)[29], within the generalized gradient approximation of the Perdew-Burke-Ernzerhof (PBE)[30] as implemented in the Vienna ab initio simulation package (VASP)[31]. The convergence criteria are determined by the energy, stress and displacement of each atom, which are set as less than $2\times10^{-3}$eV, 0.5eV/Å and $5\times10^{-3}$Å. The self-consistent force and energy convergence limit is set to be 0.002eV/Å and $10^{-6}$, while optimizing zigzag and armchair antimonene nanotubes[32]. Lengths of the Sb-Sb bonds in the optimized structures vary in a small range, while changing from nanosheets to nanotubes.

(a)

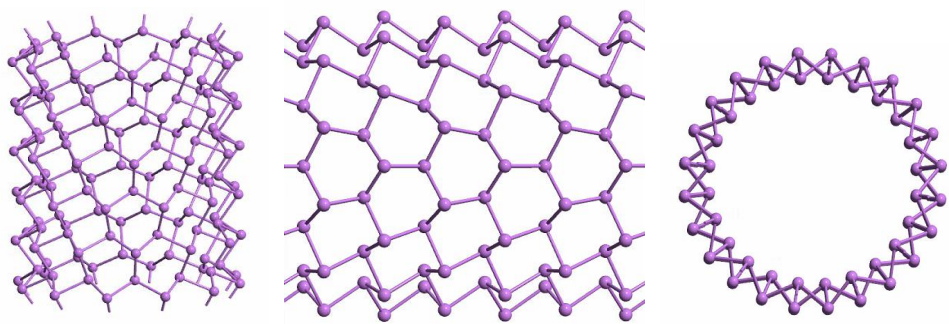

(b)

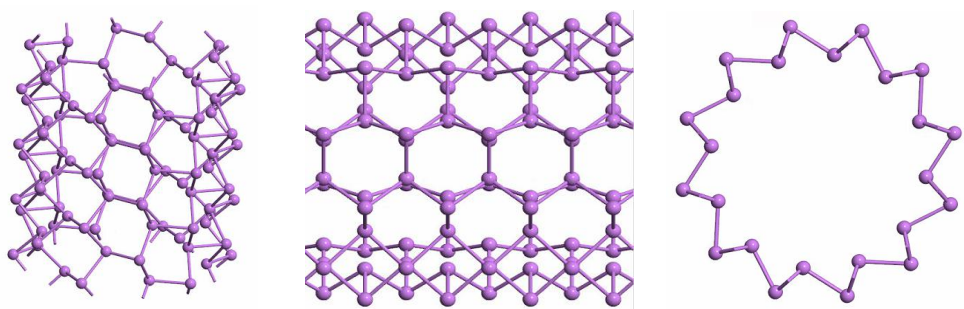

Fig.1 Topside and upside for perfect antimonene nanotubes.(a)The type of zigzag antimonene nanotube. (b)The type of armchair antimonene nanotube.

There are two possible orientations of nanotube structures that Stone-Wales defects are formed on zigzag or armchair nanotubes with a particular length and diameter. In a two-dimensional single-layered antimonene, Stone-Wales defect is defined that a pair of Sb atoms are rotated by 90° to form a topology of two five-membered and two seven-membered rings. The orientation of a Stone-Wales defect can be described by the Sb-Sb band original direction. In the antimonene sheet, each Sb atom is connected to three surrounding atoms to form a stable six-membered ring similar to graphene[33]. This results in the orientation of the Sb-Sb bonds in three different directions(shown in Fig.2 (a)(b)). When the zigzag nanotubes are formed, due to the symmetry of the nanotubes extending in the axial direction, the three orientations of the bonds may become axial($Sb_1$-$Sb_2$) and non-axial($Sb_1$-$Sb_3$,$Sb_1$-$Sb_4$). Similarly, according to the orthogonality between the armchair-type and the zigzag-type, the orientations of Stone-Wales defects are divided into perpendicular to the axial direction($Sb_1$-$Sb_2$) and intersect with the axial direction($Sb_1$-$Sb_3$,$Sb_1$-$Sb_4$). The electronic properties and structural optimization of those models are based on the same methods.

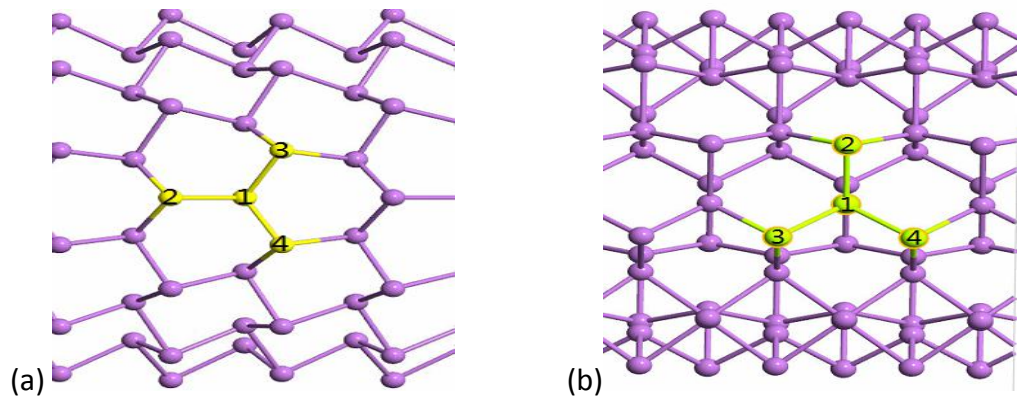

(a) (b)

Fig.2 The schematic diagram for the orientation angles of the Stone-Wales defects. (a)The type of zigzag antimonene nanotube. (b)The type of armchair antimonene nanotube.

### 3. Results and discussion

3.1 Geometric structures of Stone-Wales defective antimonene nanotubes

The optimized structures of Stone-Wales defective antimonene of zigzag and armchair nanotubes is shown in Fig.3 respectively. The antimony atoms in the antimonene are hybridized with $sp^2$ and $sp^3$ to form a stable hexagon ring structure, so that all atoms of the monolayer are not in the plane[34]. Armchair and zigzag antimonene nanotubes exhibit a low-buckling structure with highly anisotropic corrugations in which adjacent rows of Sb atoms are alternately wrinkled in the armchair and zigzag directions, respectively[35]. The presence of warped structure makes the surface of the antimonene nanotubes uneven, and it can be clearly seen that the Sb atoms are divided into two layers. In order to prevent obvious deformation of the connection between adjacent Sb atoms, the diameter of the nanotubes must be sufficiently large to avoid the interference between the inner Sb atoms. Therefore, the appropriate nanotube structures that are stable in the presence of Stone-Wales defects are used for the research. Fig.3 illustrates the Stone-Wales defective zigzag and armchair nanotubes after geometric optimization. Stone-Wales defect converts the four adjacent hexagons (6-6-6-6) into two pentagons and two heptagons (5-7-7-5). In contrast to the perfect nanotubes, the position of the atoms around the defect changes significantly, especially as the atom is closer to the defect, the greater the displacement is. Defects have a certain degree of symmetry, and the direction can be determined by the pair of Sb atoms in which the two heptagons coincide. We set the direction of the Sb-Sb bond at the center of the defect to be the orientation of the defect[36]. The orientation angles of these defects are 20.969°, 32.307°, 49.556°, and 11.815°, respectively. Due to the warped structure of the antimonene, the structure of the defect is irregular. The pentagon and hexagonal structures exhibit a spatial configuration that intermingles adjacent rows of Sb atoms alternately wrinkled, but the length of the band and the angle of the Sb-Sb band vary within a small range to maintain the stability of the whole nanotubes.

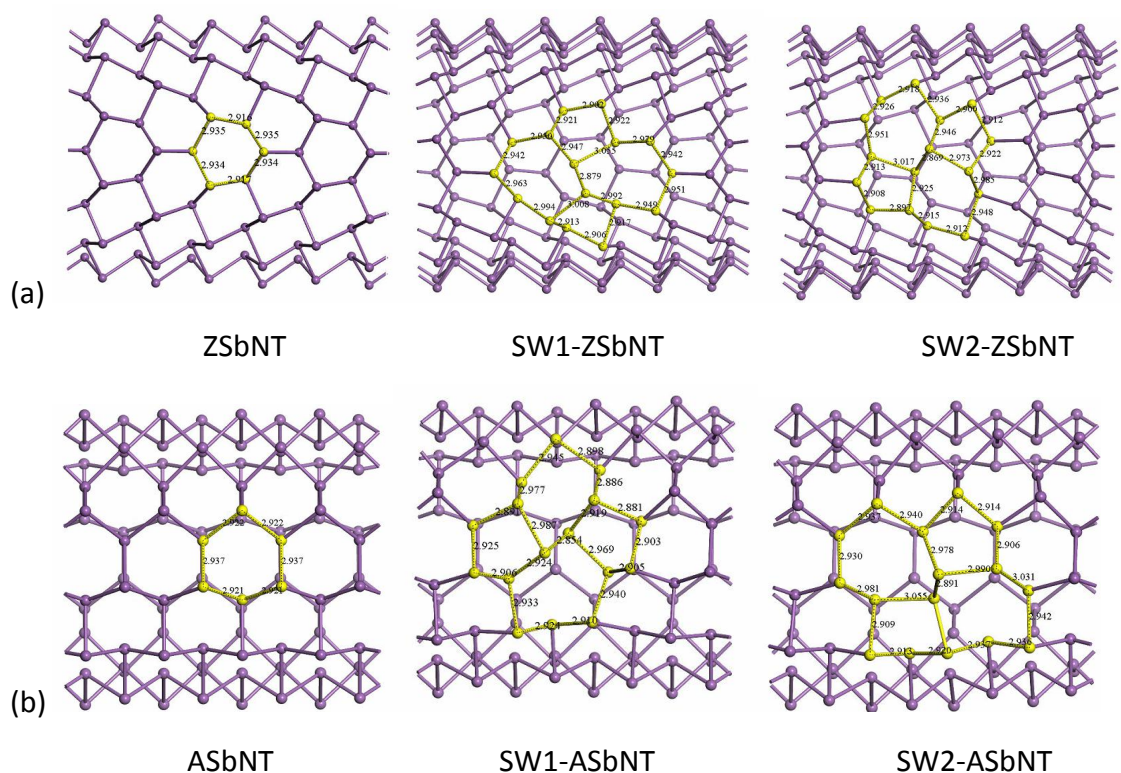

(a) ZSbNT     SW1-ZSbNT     SW2-ZSbNT

(b) ASbNT     SW1-ASbNT     SW2-ASbNT

Fig.3 The geometric structures of the perfect and Stone-Wales defective antimonene nanotubes. (a)The type of zigzag antimonene nanotube. (b)The type of armchair antimonene nanotube.

3.2 Formation and stability of Stone-Wales defect in antimonene nanotubes

According to the formation mechanism of defects, the atomic vacancies or the transfer of atomic positions may cause the atom missing or the hybridization of bonds is changed significantly, result that the total energy of the crystal structure changes dramatically. However Stone-Wales defect is relatively easy to form because it preserves the distance between the atoms and the bonding modus consistent with the former. We calculate the formation energy to study the degree of stability of the defective structure[37]. The formation energy is defined as the prerequisite energy when nanotubes form the Stone-Wales defect calculated as the discrepancy between the total energy of the perfect structure and the defect structure. The value of the formation energy($E_f$) can be estimated as $E_f=E_d-E_p$, where $E_d$ and $E_p$ are the total energy of defective and perfect nanotubes respectively. We list the formation energies for the Stone-Wales defects of ZSbNT and ASbNT in Table1. The formation energy of these structures are lower than that on graphene[38] and silicene[39], which indicates the Stone-Wales defect is easier to form for the antimonene nanotubes. The defects are sensitive to their orientations and type of nanotubes[40]. For the zigzag antimonene nanotubes, the defects of 20.969° and 32.307° have the lowest and highest formation energies. This is mainly due to the rolling-up strain of the Stone-Wales defect. The small orientation angle represents a large angle between the axis of the ZSbNT and the rotated Sb-Sb bond, which results in lower rolling-up strain and formation energy. Similar results can be found in the study of Stone-Wales defects in CNTs[41]. In contrast, for the armchair nanotubes orthogonal to ZSbNTs, the formation energy of 49.556° is larger than that of 11.815° because of the original Sb-Sb band is vertical with axial.

Table 1
Calculated structural and electronic properties for perfect and Stone-Wales defective zigzag/armchair antimonene nanotubes: Sb-Sb distance ($d_{Sb-Sb}$), formation energy ($E_f$).

|  | Diameter (Å) | Tube length(Å) | $d_{Sb-Sb}$ (Å) | Orientation angles(deg.) | Band gaps(eV) | $E_f$ (eV) |
|---|---|---|---|---|---|---|
| ZSbNT | 17.742 | 19.031 | 2.916-2.935 | 0 | 1.489 | 0 |
| SW1-ZSbNT | 17.512 | 19.048 | 2.902-3.008 | 20.969 | 1.336 | 0.582 |
| SW2-ZSbNT | 17.688 | 19.025 | 2.869-3.017 | 32.307 | 1.404 | 0.724 |
| ASbNT | 15.579 | 14.420 | 2.921-2.937 | 0 | 1.391 | 0 |
| SW1-ASbNT | 15.491 | 14.437 | 2.854-2.987 | 49.556 | 1.299 | 0.754 |
| SW2-ASbNT | 15.511 | 14.265 | 2.891-3.055 | 11.815 | 1.212 | 0.788 |

3.3 Electronic structures of the perfect and Stone-Wales defective antimonene nanotubes

The bulk structure of Sb is a typical semimetallic material, the single layer of antimonene is an indirect semiconductor and the antimonene nanotube is a direct gap semiconductor(shown in Fig.4 (a)(c))[42]. For the zigzag antimonene nanotubes, the band gaps of two nanotubes with SW defects are slightly smaller than that of the perfect nanotube. The defects of SW1-ZSbNT and SW2-ZSbNT narrow the band gap of the ZSbNT in values of 0.153eV and 0.085eV, respectively. It is noted that the conduction bands move toward low energy range because of the SW defect formation. Due to the complexity of the bonding mode for Sb atoms in the defective area, the valence bands of the defective nanotubes are more concentrated[43]. As shown in Fig.4 (a), for the perfect antimonene, the electrons in the periphery of Sb atoms form a stable structure by sp3 hybridization, in which each Sb atom forms a σ bond with three antimony atoms in the surrounding, and a pair of electrons are separated from the structure to form the π bond. The band energy near Fermi of the antimonene nanotube is consistent with Sb

5p states and 5s states. The partial density of states for various orbitals are presented, which suggests that the electrons of 5p orbitals accounted for the major contribution. This is similar to the case of the density of states of a defective nanotube(shown in Fig.4 (b)(c)). The high similarity in band structures and density of states between the pure ZSbNT and defective ZSbNTs indicates that the Stone-Wales defect has little effect on the electronic structure of the antimonene nanotube, and no significant differences can be found in their electronic properties. This indirectly confirms that the difference of the total energy between perfect and defect structures is small and the length and angle of Sb-Sb bonds are not significantly changed when the defect is formed(shown in Table1). With the same setting for calculating the properties of the ZSbNTs, the electronic properties of the perfect and defective armchair antimonene nanotubes are calculated(shown in Fig.4 (d)(e)(f)). The same conclusion can be obtained in the nanotubes for armchair.

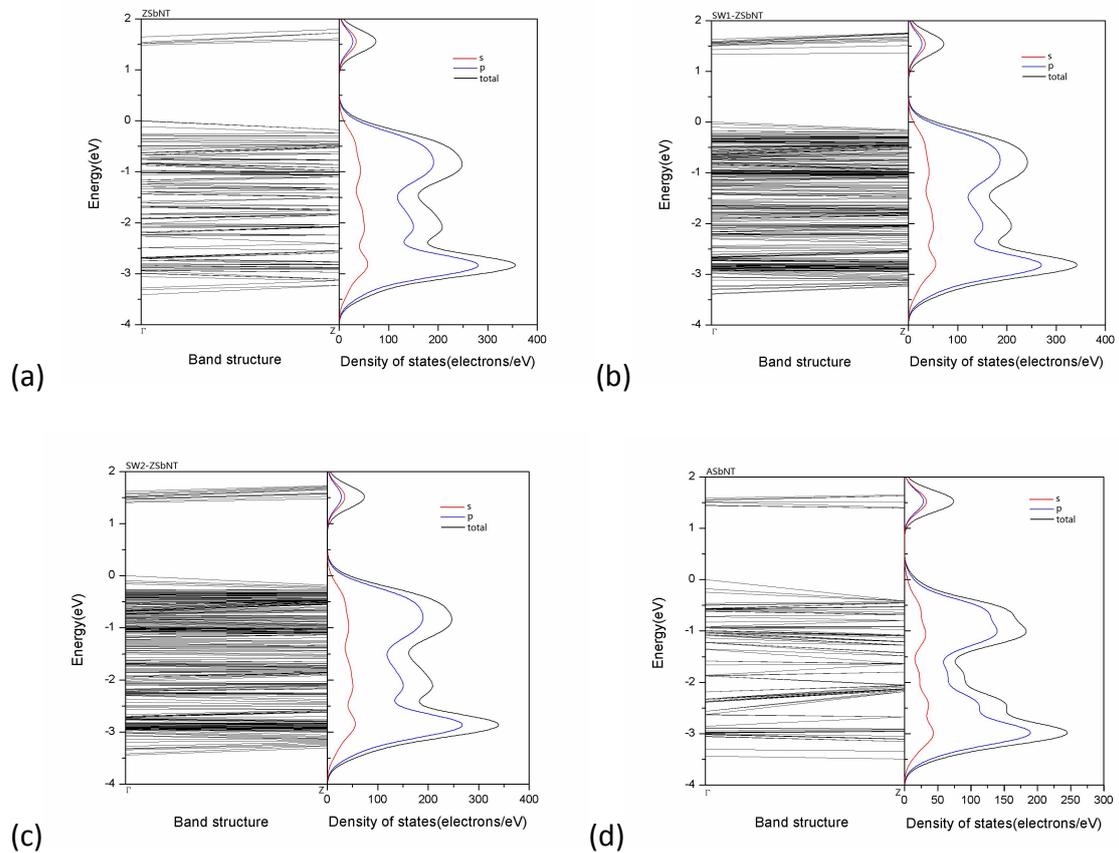

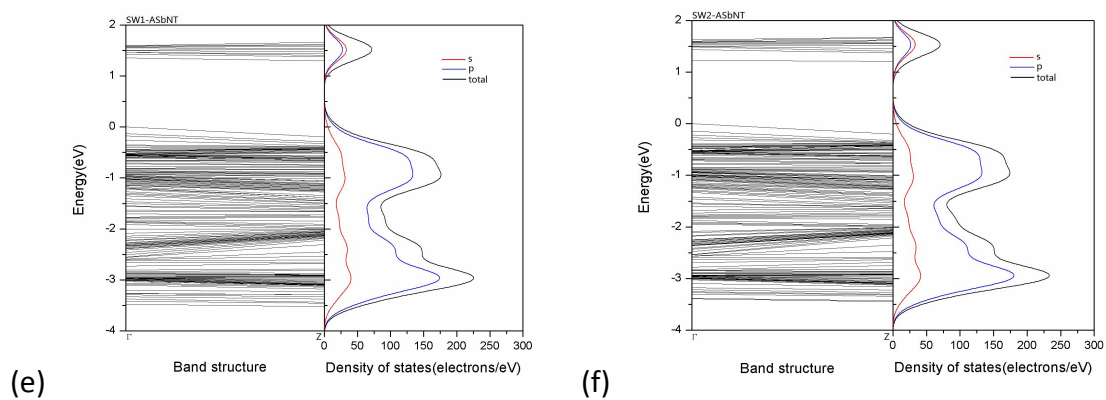

(e)                                                  (f)

Fig.4 The electronic properties of antimonene nanotubes. (a)(b)(c)The band structures and partial density of states of ZSbNTs. (d)(e)(f)The band structures and partial density of states of ASbNTs.

To further investigate the properties of the defective structures, electron density isosurfaces near Fermi are calculated(shown in Fig.5)[44]. The defect energy level near Fermi surface comprises the valence band maximum(VBM) and the conduction band minimum (CBM), concentrated on a pair of Sb atoms in the center of defective region. The p states of atomic orbitals contribute more to the highest-energy valence band and lowest-energy conduction band.

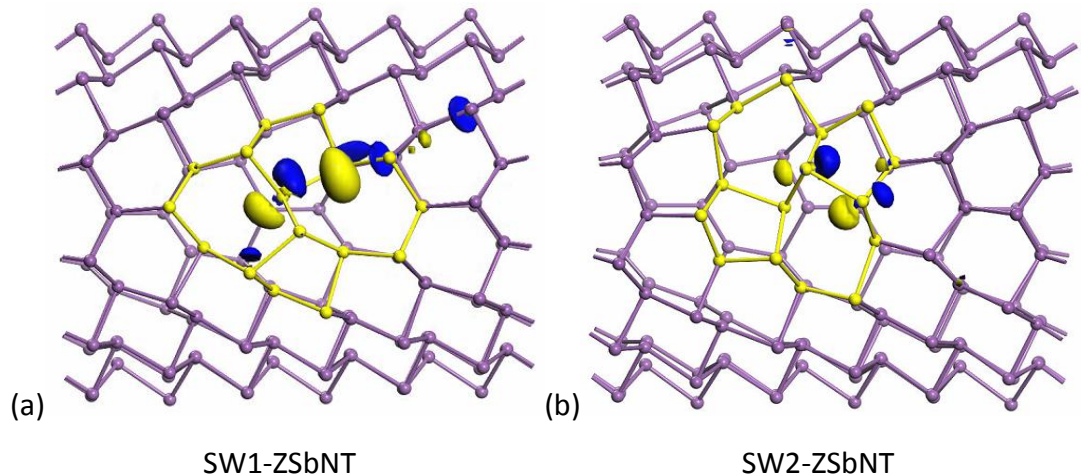

(a)                                             (b)

         SW1-ZSbNT                             SW2-ZSbNT

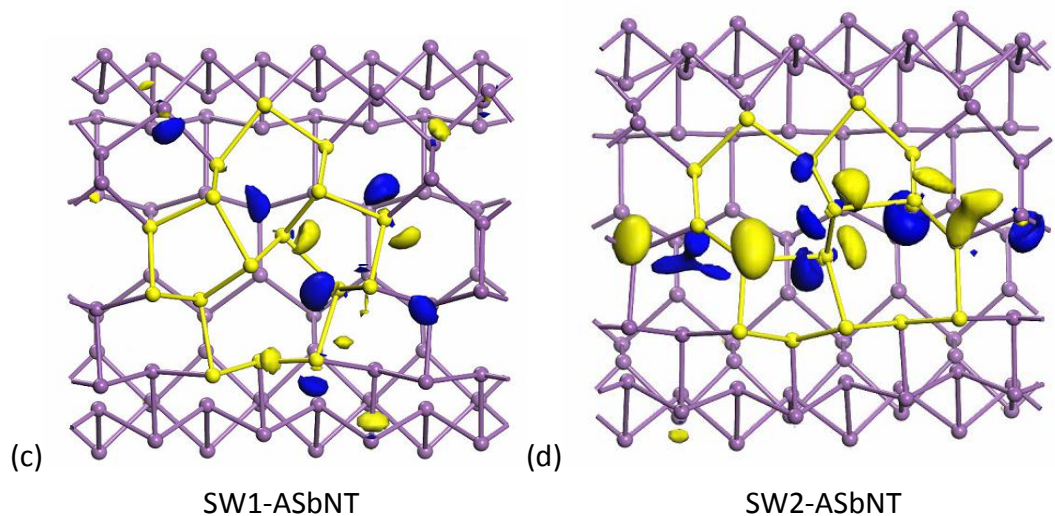

(c) SW1-ASbNT  (d) SW2-ASbNT

Fig.5 Electron density isosurfaces near Fermi of defective antimonene nanotubes. (a)(b)The electron density isosurface of ZSbNTs. (c)(d)The electron density isosurface of ASbNTs.

## 4. Conclusions

The first-principle calculations based on density functional theory are perfumed to study the geometric structures and electronic properties of Stone-Wales defective antimonene nanotubes. Stone-Wales defect converts the four adjacent hexagons into two pentagons and two heptagons. Four nanotubes are investigated according to the possible orientations of the defects. The formation energies and structures of banding suggest the stability of Stone-Wales defect in antimonene nanotubes. We investigate the mechanism with defects forming by calculating band structures, density of states and atomic orbitals. The antimonene nanotube is a direct gap semiconductor and the influence of defects on electronic properties are inconspicuous. Our researches reveal the characteristics of antimonene nanotube with defect forming, which are significant for the applications of low-dimension nanostructures of antimony on nanoelectronic devices.

## 5. Acknowledgments

The authors would like to acknowledge the support by 2018 China National College students' Innovative and Entrepreneurial Training Program Funding Projects.